\documentclass[two column, prA]{revtex4}%
\usepackage{amsmath}
\usepackage{graphicx}
\usepackage{amsfonts}
\usepackage{amssymb}
\usepackage{color}%
\providecommand{\U}[1]{\protect\rule{.1in}{.1in}}
\ifx\pdfoutput\relax\let\pdfoutput=\undefined\fi
\newcount\msipdfoutput
\ifx\pdfoutput\undefined\else
\ifcase\pdfoutput\else
\ifx\paperwidth\undefined\else
\ifdim\paperheight=0pt\relax\else\pdfpageheight\paperheight\fi
\ifdim\paperwidth=0pt\relax\else\pdfpagewidth\paperwidth\fi
\fi\fi\fi
\begin{document}
\title{Photon quantum mechanics in real Hilbert space}
\author{Margaret Hawton}
\affiliation{Department of Physics, Lakehead University, Thunder Bay, ON, Canada, P7B 5E1}

\begin{abstract}
Classically, electromagnetic pulses are described by real fields that couple
to charged matter and propagate causally. We will show here that real fields
of the form used in standard classical electromagnetic theory have a quantum
mechanical interpretation in which the probability density for a photon to be
at $\mathbf{x}$ is positive definite and operators representing all of the
standard physical observables exist. A covariant alternative to the
$\omega_{k}^{1/2}$ dependence that appears in most (but not all) presentations
of quantum optics and quantum field theory is presented and real Lorentz
scalar one photon advanced and retarded potentials are derived.

\end{abstract}
\maketitle

\section{\qquad Introduction}

This work was originally motivated by mathematical proofs that a positive
frequency field cannot be confined to a finite region of space. According to
the Hegerfeldt theorem a positive frequency field localized in a finite region
for an instant spreads immediately throughout space \cite{Hegerfeldt}. It has
been shown explicitly in the case of one dimensional square wells
\cite{Karpov} and for three dimensional position eigenvectors
\cite{HawtonDebierre,MaxwellQM} that this instantaneous localization is only
apparent since it is due to destructive interference of intrinsically nonlocal
counterpropagating waves. In algebraic quantum field theory (QFT),\ the
Reeh-Schleider theorem states that there are no local annihilation or creation
operators \cite{ReehSchlieder}. However, confinement of real fields to a
finite region is not a problem in classical electromagnetism (EM). It will be
proved here that this use of real fields can be extended to photon quantum
mechanics (QM).

The QM of electrons and other fermions is well understood but a consistent
first quantized theory of the photon has been elusive. Photons have two
properties not shared with electrons that have made derivation of photon QM
difficult - they are neutral and massless. While fields describing charged
particles are intrinsically complex, neutral particles should be described by
real fields. Reality of the photon wave function ensures that photons and
antiphotons, being indistinguishable, are equally probable and that, after
second quantization, their field operators become Hermitian. This property is
problematic in a first quantized theory since the standard relativistic scalar
product is zero for neutral particles. Also, while the Wigner little group
describing massive particles is the set of spatial rotations, the Wigner
little group for massless particles is cylindrically symmetrical. In their
seminal paper titled "Localized states of elementary systems", Newton and
Wigner assumed invariance under spherically symmetrical rotations and
concluded that "for equations with zero mass .. with spin 1 (i.e. Maxwell's
equations) we found that no localized states in the above sense exist. This is
an unsatisfactory .. feature of our work" \cite{NW}. Both of these
difficulties have been overcome
\cite{HawtonPosOp,HawtonDebierre,MostafazadehZamani,BabaeiMostafazadeh,MaxwellQM,WignerLittleGroup}
but here we will extend this work by formulating photon QM in terms of
physically correct real fields.

In field theory, particles that transform into themselves are represented by
real fields and Hermitian field operators \cite{GellMann}. These real fields
can be written as linear combinations of the positive frequency particle
terms, $A^{+}$, and negative frequency antiparticle terms, $A^{-}$. The real
fields $A_{c}=\left(  A_{c}^{+}+A_{c}^{-}\right)  /\sqrt{2}$ and
$A_{s}=\left(  A_{s}^{+}-A_{s}^{-}\right)  /\sqrt{2}i$ are even and odd
respectively under particle/antiparticle exchange (charge conjugation) where
the subscripts $c$ and $s$ denote Fourier cosine and sine series respectively.
The complex function defined as $A=A_{c}+iA_{s}$ describes both of these
potentials, so it will be used here to simplify the mathematics.

In Section II the photon equations of motion will be derived from the standard
Lagrangian. The fields will be assumed to be real but for mathematical
convenience these real fields will be identified with the real and imaginary
parts of a complex field. The zeroth component of the modified standard
relativistic four-current will be interpreted as a positive definite photon
number density and used to define a scalar product. Interaction with polarized
matter will be included so that these equations can be applied to photon
propagation in a nonabsorptive medium and emission and absorption of photons
by localized sources and sinks. In Section III the real Hilbert space will be
defined and operators for all the usual physical observables, including
position, will be reviewed to provide a complete first quantized description
of single-photon states. In Section IV the fields will be second quantized and
in the final Section we will summarize and conclude.

\section{Real and complex fields, Lagrangian and scalar product}

In this Section four-vector notation and complex fields whose real and
imaginary parts are even and odd under QFT charge conjugation will be defined.
The photon equations of motion, four-current and positive definite number
density will be derived from the real standard Lagrangian written in terms of
complex fields. A scalar product will be defined in position space and Fourier
transformed to momentum space to complete the Hilbert space.

Relativistic notation and SI units will be used. The contravariant space-time,
wave vector and momentum four-vectors are $x=x^{\mu}=\left(  ct,\bm{x}\right)
,$ $k=\left(  \omega_{k}/c,\mathbf{k}\right)  $ and $p=\hbar k$ where
$kx=\omega_{k}t-\mathbf{k\cdot x}$, the four-gradient is $\partial=\left(
\partial_{ct},-\mathbf{\nabla}\right)  $, the four-potential is $A\left(
t,\mathbf{x}\right)  =A^{\mu}=\left(  \frac{\phi}{c},\mathbf{A}\right)  $ or
$a\left(  t,\mathbf{k}\right)  =\left(  a_{0},\mathbf{a}\right)  $ and
$a_{\lambda}\left(  \mathbf{k}\right)  $ will denote a Lorentz invariant
scalar describing a state with definite helicity, $\lambda$. The covariant
four-vector corresponding to $U^{\mu}=\left(  U_{0},\mathbf{U}\right)  $ is
$U_{\mu}=g_{\mu\nu}U^{\nu}=\left(  U_{0},-\mathbf{U}\right)  $ where
$g_{\mu\nu}=g^{\mu\nu}$ is a $4\times4$ diagonal matrix with diagonal $\left(
1,-1,-1,-1\right)  $.

Positive and negative frequency four-potentials describing photons and
antiphotons respectively will be defined as%
\begin{align}
A_{r}^{+}\left(  x\right)   &  =\sqrt{\frac{\hbar}{\epsilon_{0}}}\int_{t}%
\frac{d\mathbf{k}}{\left(  2\pi\right)  ^{3}\omega_{k}}a_{rx^{\prime}}\left(
\mathbf{k}\right)  e^{-ikx},\label{A+}\\
A_{r}^{-}\left(  x\right)   &  =A_{r}^{+\ast}\left(  x\right)  ,\label{A-}\\
a_{rx^{\prime}}\left(  \mathbf{k}\right)   &  =a_{r}\left(  \mathbf{k}\right)
e^{ikx^{\prime}} \label{arx}%
\end{align}
for $r=c$ and $s$ where $a_{c}\left(  \mathbf{k}\right)  $ and $a_{s}\left(
\mathbf{k}\right)  $ are real, $x^{\prime}$ is a shift in the origin of the
space-time coordinates, the superscript $\ast$ denotes complex conjugation,
the subscript $t$ on the integral denotes evaluation at a fixed time $t$ and
$d\mathbf{k}\equiv d^{3}k$ is an infinitesimal volume in $\mathbf{k}$-space.
The above form was selected because lim$_{V\rightarrow\infty}\Delta
\mathbf{n}/V=d\mathbf{k}/\left(  2\pi\right)  ^{3}$ where $\Delta\mathbf{n}$
is the number of states and $\int d^{4}k\delta\left(  \omega_{k}^{2}%
/c^{2}-\left\vert \mathbf{k}\right\vert ^{2}\right)  =\int_{t}\frac
{d\mathbf{k}}{\left(  2\pi\right)  ^{3}2\omega_{k}/c}$. Since $kx$ and
$\int_{t}\frac{d\mathbf{k}}{\left(  2\pi\right)  ^{3}\omega_{k}}$ are
invariants, if $A_{r}\left(  x\right)  $ is a four-vector, then $a_{r}\left(
\mathbf{k}\right)  $ are four-vectors. The real potentials%
\begin{align}
A_{c}\left(  x\right)   &  =\frac{A_{c}^{+}\left(  x\right)  +A_{c}^{-}\left(
x\right)  }{\sqrt{2}}=\sqrt{2}\operatorname{Re}\left(  A_{c}^{+}\right)
\label{Ac+}\\
&  =\sqrt{\frac{2\hbar}{\epsilon_{0}}}\int_{t}\frac{d\mathbf{k}}{\left(
2\pi\right)  ^{3}\omega_{k}}a_{c}\left(  \mathbf{k}\right)  \cos\left[
k\left(  x-x^{\prime}\right)  \right]  ,\label{Ac}\\
A_{s}\left(  x\right)   &  =\frac{A_{s}^{+}\left(  x\right)  -A_{s}^{-}\left(
x\right)  }{\sqrt{2}i}=\sqrt{2}\operatorname{Im}\left(  A_{s}^{+}\right)
\label{As+}\\
&  =-\sqrt{\frac{2\hbar}{\epsilon_{0}}}\int_{t}\frac{d\mathbf{k}}{\left(
2\pi\right)  ^{3}\omega_{k}}a_{s}\left(  \mathbf{k}\right)  \sin\left[
k\left(  x-x^{\prime}\right)  \right]  , \label{As}%
\end{align}
are even and odd respectively under spacetime reflection. The factor $\sqrt
{2}$ normalizes the cosine and sine functions squared whose average over a
period is $\frac{1}{2}$. For CPT symmetric particles this is equivalent to QFT
charge conjugation. The functions $A_{c}\left(  x\right)  $ and $A_{s}\left(
x\right)  $ are real by definition. The complex four-potential will be defined
as%
\begin{equation}
A\left(  x\right)  =A_{c}\left(  x\right)  +iA_{s}\left(  x\right)  .
\label{Acs}%
\end{equation}
Substitution of (\ref{Ac}) and (\ref{As}) in (\ref{Acs}) gives%
\begin{align}
A\left(  x\right)   &  =\sqrt{\frac{\hbar}{2\epsilon_{0}}}\int_{t}%
\frac{d\mathbf{k}}{\left(  2\pi\right)  ^{3}\omega_{k}}\left[  \left(
a_{c}\left(  \mathbf{k}\right)  -a_{s}\left(  \mathbf{k}\right)  \right)
e^{ik\left(  x-x^{\prime}\right)  }\right. \nonumber\\
&  \left.  +\left(  a_{c}\left(  \mathbf{k}\right)  +a_{s}\left(
\mathbf{k}\right)  \right)  e^{-ik\left(  x-x^{\prime}\right)  }\right]
\label{A+-}%
\end{align}
which shows that $A$ is positive frequency only if $a_{c}=a_{s}$. The sines
and cosines or the positive and negative frequency plane waves $e^{-ikx}$ and
$e^{ikx}$ form a basis, but here we start from the premise that only real
fields are physically correct and only odd real fields couple correctly to
charged matter. The positive and negative frequency fields are introduced to
simplify the mathematics. The complex electric and magnetic fields are
$\mathbf{E}=-\partial_{t}\mathbf{A}-\mathbf{\nabla}\phi$ and $\mathbf{B}%
=\mathbf{\nabla}\times\mathbf{A}$ and the antisymmetric Faraday tensor is
\begin{equation}
\mathcal{F^{\mu\nu}=}\partial^{\mu}A^{\nu}-\partial^{\nu}A^{\mu}
\label{Ftensor}%
\end{equation}
where $\mathcal{F}^{00}=\mathcal{F}^{ii}=0$, $\mathcal{F}^{i0}=-\mathcal{F}%
^{0i}=E_{i}/c$, $\mathcal{F}^{ij}=-\mathcal{F}^{ji}=-\epsilon_{ijk}B_{k}$ and
$\epsilon_{ijk}$ is the Levi-Civita symbol.

The Lagrangian describing two real fields can be written in complex form
provided this field and its complex conjugate are treated as formally
independent \cite{CT}. In the presence of a matter four-current density
$J_{m}^{\nu}+c.c,$ the real Lagrangian density will be written as%
\begin{align}
\mathcal{L}  &  =\mathcal{L}_{std}+\mathcal{L}_{int}\label{L}\\
\mathcal{L}_{std}  &  =\epsilon_{0}\left(  \mathbf{E\cdot E}^{\ast}%
-c^{2}\mathbf{B\cdot B}^{\ast}\right) \label{Lstd}\\
&  =-\frac{1}{2}\epsilon_{0}c^{2}\mathcal{F^{\mu\nu}F_{\mu\nu}^{\ast}%
}\nonumber\\
\mathcal{L}_{int}  &  =-J_{m}^{\nu\ast}A_{\nu}-J_{m}^{\nu}A_{\nu}^{\ast}
\label{Lint}%
\end{align}
where $c$ is the speed of light, $\epsilon_{0}$ is the dielectric
permittivity, $J_{m}^{\mu}=J_{cm}^{\mu}+iJ_{sm}^{\mu}$ and \thinspace
$J_{m}^{\mu}=\left(  \rho_{m}c,\mathbf{J}_{m}\right)  $. The Lagrange
equations of motion are then
\begin{equation}
\partial_{\mu}\left[  \frac{\partial\mathcal{L}}{\partial\left(  \partial
_{\mu}A_{\nu}^{^{\ast}}\right)  }\right]  =\frac{\partial\mathcal{L}}{\partial
A_{\nu}^{^{\ast}}} \label{Lagrange}%
\end{equation}
where the momentum conjugate to $A_{\nu}^{^{\ast}}$ is%
\begin{equation}
\Pi^{\mu\nu}\equiv\frac{\partial\mathcal{L}}{\partial\left(  \partial_{\mu
}A_{\nu}^{^{\ast}}\right)  }=-\epsilon_{0}c^{2}\mathcal{F}^{\mu\nu}.
\label{Pi}%
\end{equation}
Eq. (\ref{Lagrange}) then gives
\begin{equation}
\epsilon_{0}c^{2}\partial_{\mu}\mathcal{F^{\mu\nu}}=J_{m}^{\nu}, \label{Leqn}%
\end{equation}
that can be written as the Maxwell equations (ME)%
\begin{equation}
\mathbf{\nabla}\cdot\mathbf{D}=\rho_{m},\ \ \mathbf{\nabla}\times
\mathbf{H-}\partial_{t}\mathbf{D}=\mathbf{J}_{m} \label{MEs}%
\end{equation}
where $\mathbf{D}=\epsilon_{0}\mathbf{E}$, $\mathbf{B}=\mu_{0}\mathbf{H}$ and
$\mu_{0}\epsilon_{0}=1/c^{2}$. Substitution of (\ref{Ftensor}) in (\ref{Leqn})
gives the wave equation
\begin{equation}
\partial_{\mu}\partial^{\mu}A^{\nu}=\partial^{\nu}\Lambda+\mu_{0}J_{m}^{\nu}
\label{Awave}%
\end{equation}
where
\begin{equation}
\partial_{\mu}\partial^{\mu}=\partial_{ct}^{2}-\mathbf{\nabla}^{2}%
\equiv\square. \label{del2}%
\end{equation}
The gauge is determined by
\begin{equation}
\Lambda=\partial_{\mu}A^{\mu}\mathbf{.} \label{Lambda}%
\end{equation}
In the Coulomb gauge $\nabla\cdot\mathbf{A}=0$, so there are no longitudinal
modes and the scalar field satisfying $\nabla^{2}\phi=-\rho/\epsilon_{0}$
responds instantaneously to changes in charge density. Only the transverse
modes propagate at the speed of light and are second quantized in quantum
electrodynamics (QED) to allow creation and annihilation of physical photons.
In the Lorenz gauge $\Lambda=0$ inserted into (\ref{Awave}) gives
$\partial_{\mu}\partial^{\mu}A^{\nu}=\mu_{0}J_{m}^{\nu}$. In this gauge all
four components of $A$ describing the scalar, longitudinal and transverse
photon modes propagate at the speed of light and are second quantized in QED.
Each complex equation in this paragraph is equivalent to two real equations;
one for the even potentials and one for the odd potentials.

Using these complex fields a positive definite photon number density and a
scalar product can be derived starting with the global phase change
$A\longrightarrow e^{i\alpha}A$, $A^{\ast}\longrightarrow e^{-i\alpha}A^{\ast
}$ \ that is a symmetry of the free space Lagrangian $\mathcal{L}_{std}$. For
an infinitesimal change in $A$, $\delta A\simeq i\alpha A$ and $\delta
A^{\ast}\simeq-i\alpha A^{\ast}$ so, using $\Pi^{\mu\nu}=-\epsilon_{0}%
c^{2}\mathcal{F}^{\mu\nu}$, the Noether four-current density becomes $J^{\mu
}\left(  x\right)  \propto\mathcal{F}^{\mu\nu\ast}A_{\nu}-\mathcal{F}^{\mu\nu
}A_{\nu}^{\ast}$. The four-current density $J^{\mu}\left(  x\right)
\propto\mathcal{F}^{\mu\nu}\left(  x\right)  A_{\nu}\left(  x\right)  $ was
first obtained in \cite{HawtonMelde} where it was used to derive a Hermitian
number density operator.

Based on this expression negative frequency waves make a negative contribution
to the number density $J^{0}\left(  x\right)  $. To solve this problem
Mostafazadeh and coworkers defined the sign of frequency operator
$\widehat{\epsilon}$ as \cite{MostafazadehZamani,BabaeiMostafazadeh}
\begin{equation}
\widehat{\epsilon}\equiv i\left(  -\nabla^{2}\right)  ^{-1/2}\partial_{ct}.
\label{epsilon}%
\end{equation}
such that, if $A$ satisfies the photon wave equation, then $\widehat{\epsilon
}A$ also satisfies the photon wave equation. With%
\begin{equation}
\widetilde{\mathcal{F}}^{\mu\nu}\equiv\widehat{\epsilon}\mathcal{F}^{\mu\nu}
\label{Ftilde}%
\end{equation}
it can be verified by substitution that
\begin{equation}
J^{\mu}\left(  x\right)  =\frac{\epsilon_{0}c^{2}}{2\hbar}%
\widetilde{\mathcal{F}}^{\mu\nu\ast}A_{\nu}+c.c. \label{J+ve}%
\end{equation}
satisfies a continuity equation in the homogenous case $J_{m}=0$ and that
$J^{0}\left(  x\right)  $ is positive definite for both positive and negative
frequency fields. The mathematics of the operator $\widehat{\epsilon}$ is
discussed in \cite{MostafazadehZamani,BabaeiMostafazadeh}. The operator
$\left(  -\nabla^{2}\right)  ^{-1/2}$ just extracts a factor $\left\vert
\mathbf{k}\right\vert ^{-1/2}$ from the plane wave $e^{-i\epsilon kx}$, while
$i\partial^{\mu}e^{-i\epsilon k^{\nu}x_{\nu}}=\epsilon k^{\mu}e^{-i\epsilon
k^{\nu}x_{\nu}}$ so that the operator $\widehat{\epsilon}$ extracts the sign
of frequency, $\epsilon$. Thus $\widehat{\epsilon}\partial^{\mu}\cos\left(
kx\right)  =k^{\mu}\cos\left(  kx\right)  $ and $\widehat{\epsilon}%
\partial^{\mu}\sin\left(  kx\right)  =k^{\mu}\sin\left(  kx\right)  $ gives%
\begin{equation}
J^{\mu}\left(  x\right)  =\frac{\epsilon_{0}c^{2}}{\hbar}\left(
\widetilde{\mathcal{F}}_{c}^{\mu\nu}A_{c\nu}+\widetilde{\mathcal{F}}_{s}%
^{\mu\nu}A_{s\nu}\right)  \label{Jcs}%
\end{equation}
where $\widetilde{\mathcal{F}}_{c}^{\mu\nu}$ and $\widetilde{\mathcal{F}}%
_{s}^{\mu\nu}$ are, respectively, cosine and sine series. If $J_{m}=0$, $J$
satisfies a continuity equation and the spatial integral of the number density
$J^{0}\left(  x\right)  $ is conserved.

The four-current (\ref{J+ve}) or (\ref{Jcs}) is not gauge invariant due to its
dependence on $A$. In the Coulomb gauge only transverse waves propagate, while
in the Lorenz gauge longitudinal and transverse photons exist, but their
contributions to the scalar product cancel in free space \cite{MaxwellQM}.
With the mutually orthogonal polarization unit vectors $e^{\mu}$ defined such
that 0 is time-like, 1 and 2 are transverse and 3 is longitudinal,
$e_{0}=n^{\mu}=\left(  1,0,0,0\right)  ,\ \mathbf{e}_{3}\left(  \mathbf{k}%
\right)  =\mathbf{e}_{\mathbf{k}}=\mathbf{k}/\left\vert \mathbf{k}\right\vert
$ and the definite helicity transverse unit vectors are%
\begin{equation}
\mathbf{e}_{\lambda}=\frac{1}{\sqrt{2}}\left(  \mathbf{e}_{\theta}%
+i\lambda\mathbf{e}_{\phi}\right)  \label{transverseeigenvectors}%
\end{equation}
for $\lambda=\pm1\ $where $\mathbf{e}_{\theta},$ $\mathbf{e}_{\phi}$ and
$\mathbf{e}_{\mathbf{k}}$ are $\mathbf{k}$-space spherical polar unit vectors
on the $t$-hyperplane. Since $e^{\mu}$ is a four-vector its coefficient in any
expression for $a^{\mu}$ should be a Lorentz invariant scalar. Writing
$\widetilde{\mathcal{F}}^{0\nu}$ as an electric field divided by $c$ and
including only transverse photons with helicity $\lambda=\pm1$, the scalar
product at a fixed time $t$ will be defined as%
\begin{equation}
\left(  A_{1},A_{2}\right)  _{t}=\frac{\epsilon_{0}}{\hbar}\sum_{\lambda=\pm
1}\int_{t}d\mathbf{x}\widetilde{\mathbf{E}}_{1\lambda}^{\ast}\left(  x\right)
\cdot\mathbf{A}_{2\lambda}\left(  x\right)  \label{EAproduct}%
\end{equation}
where the transverse $\lambda=\pm1$ components of the potentials (\ref{Acs}),
(\ref{Ac}) and (\ref{As}) and their dual fields are%
\begin{align}
\mathbf{A}_{\lambda}\left(  x\right)   &  =\mathbf{A}_{c\lambda}\left(
x\right)  +i\mathbf{A}_{s\lambda}\left(  x\right)  ,\label{Al}\\
\mathbf{A}_{c\lambda}\left(  x\right)   &  =\sqrt{\frac{2\hbar}{\epsilon_{0}}%
}\int_{t}\frac{d\mathbf{k}}{\left(  2\pi\right)  ^{3}\omega_{k}}%
\mathbf{a}_{c\lambda}\left(  \mathbf{k}\right)  \cos\left[  k\left(
x-x^{\prime}\right)  \right]  ,\label{Acl}\\
\mathbf{A}_{s\lambda}\left(  x\right)   &  =-\sqrt{\frac{2\hbar}{\epsilon_{0}%
}}\int_{t}\frac{d\mathbf{k}}{\left(  2\pi\right)  ^{3}\omega_{k}}%
\mathbf{a}_{s\lambda}\left(  \mathbf{k}\right)  \sin\left[  k\left(
x-x^{\prime}\right)  \right]  ,\label{Asl}\\
\widetilde{\mathbf{E}}_{r\lambda}\left(  x\right)   &  =\widehat{\epsilon
}\mathbf{E}_{r\lambda}\left(  x\right)  =-\widehat{\epsilon}\partial
_{t}\mathbf{A}_{r\lambda}\left(  x\right)  . \label{Edual}%
\end{align}
In these expressions%
\begin{equation}
\mathbf{a}_{rx^{\prime}\lambda}\left(  \mathbf{k}\right)  \equiv
a_{rx^{\prime}\lambda}\left(  \mathbf{k}\right)  \mathbf{e}_{\lambda}\left(
\mathbf{k}\right)  ,\ a_{rx^{\prime}\lambda}\left(  \mathbf{k}\right)
=a_{r\lambda}\left(  \mathbf{k}\right)  e^{ikx^{\prime}} \label{alambda}%
\end{equation}
where $a_{r\lambda}\left(  \mathbf{k}\right)  $ is a Lorentz invariant scalar.
Differentation of%
\begin{equation}
\mathbf{A}_{r\lambda}^{+}\left(  x\right)  =\sqrt{\frac{\hbar}{\epsilon_{0}}%
}\int_{t}\frac{d\mathbf{k}}{\left(  2\pi\right)  ^{3}\omega_{k}}%
\mathbf{a}_{rx^{\prime}\lambda}\left(  \mathbf{k}\right)  e^{-ikx} \label{Ar+}%
\end{equation}
gives the positive frequency dual electric fields%
\begin{equation}
\widetilde{\mathbf{E}}_{r\lambda}^{+}\left(  x\right)  =\sqrt{\frac{\hbar
}{\epsilon_{0}}}\int_{t}\frac{d\mathbf{k}}{\left(  2\pi\right)  ^{3}%
}\mathbf{a}_{rx^{\prime}\lambda}\left(  \mathbf{k}\right)  e^{-ikx}.
\label{Er+}%
\end{equation}
As in (\ref{Ac+}) to (\ref{As}) $\mathbf{A}_{c\lambda}=\sqrt{2}%
\operatorname{Re}\mathbf{A}_{c\lambda}^{+}$, $\mathbf{A}_{s\lambda}=\sqrt
{2}\operatorname{Im}\mathbf{A}_{s\lambda}^{+}$, $\widetilde{\mathbf{E}%
}_{c\lambda}=\sqrt{2}\operatorname{Re}\widetilde{\mathbf{E}}_{c\lambda}^{+}$
and $\widetilde{\mathbf{E}}_{s\lambda}=\sqrt{2}\operatorname{Im}%
\widetilde{\mathbf{E}}_{s\lambda}^{+}$. Substitution of (\ref{Al}) gives the
scalar product (\ref{EAproduct}) in terms of its real even and odd components
as%
\begin{align}
\left(  A_{1},A_{2}\right)  _{t}  &  =\frac{\epsilon_{0}}{\hbar}\sum
_{\lambda=\pm1}\int_{t}d\mathbf{x}\left[  \widetilde{\mathbf{E}}_{1c\lambda
}\left(  x\right)  \cdot\mathbf{A}_{2c\lambda}\left(  x\right)  \right.
\nonumber\\
&  \left.  +\widetilde{\mathbf{E}}_{1s\lambda}\left(  x\right)  \cdot
\mathbf{A}_{2s\lambda}\left(  x\right)  \right]  . \label{EAagain}%
\end{align}
Bra-ket notation will be defined as in Schr\"{o}dinger QM so that
(\ref{EAagain}) can be written as
\begin{equation}
\left(  A_{1},A_{2}\right)  _{t}=\frac{\epsilon_{0}}{\hbar}\sum_{\lambda=\pm
1}\left[  \left\langle \widetilde{\mathbf{E}}_{1c\lambda}\cdot\mathbf{A}%
_{2c\lambda}\right\rangle +\left\langle \widetilde{\mathbf{E}}_{1s\lambda
}\cdot\mathbf{A}_{2s\lambda}\right\rangle \right]  \label{EAcs}%
\end{equation}
where%
\begin{align}
\left\langle \widetilde{\mathbf{E}}_{1r\lambda}\cdot\mathbf{A}_{2r\lambda
}\right\rangle  &  =\sum_{j=1}^{3}\left\langle \widetilde{E}_{1r\lambda
j}|A_{2c\lambda j}\right\rangle \label{BraKet}\\
&  =\int d\mathbf{x}\widetilde{\mathbf{E}}_{1r\lambda}\left(  x\right)
\cdot\mathbf{A}_{2r\lambda}\left(  x\right) \nonumber\\
&  =\int\frac{d\mathbf{k}}{\left(  2\pi\right)  ^{3}}a_{1r\lambda}\left(
\mathbf{k}\right)  \frac{a_{2r\lambda}\left(  \mathbf{k}\right)  }{\omega_{k}%
}e^{ik\left(  x_{1}-x_{2}\right)  }.\nonumber
\end{align}
Equality of these last two expressions is an expression of the
Parseval-Plancherel identity. By inspection of (\ref{Er+}) and (\ref{Ar+}),
$\mathbf{a}_{rx^{\prime}\lambda}\left(  \mathbf{k}\right)  $ are Fourier
transforms of $\sqrt{\frac{\epsilon_{0}}{\hbar}}\widetilde{\mathbf{E}%
}_{r\lambda}^{+}\left(  x\right)  $ while $\mathbf{a}_{rx^{\prime}\lambda
}\left(  \mathbf{k}\right)  /\omega_{k}$ are Fourier transforms of
$\sqrt{\frac{\epsilon_{0}}{\hbar}}\mathbf{A}_{r\lambda}^{+}\left(  x\right)
$. Explicitly, in $\mathbf{k}$-space, the scalar product is
\begin{align}
\left(  A_{1},A_{2}\right)  _{t}  &  =\sum_{\lambda=\pm1}\int_{t}%
\frac{d\mathbf{k}}{\left(  2\pi\right)  ^{3}\omega_{k}}\left[  a_{1c\lambda
}\left(  \mathbf{k}\right)  a_{2c\lambda}\left(  \mathbf{k}\right)  \right.
\nonumber\\
&  \left.  +a_{1s\lambda}\left(  \mathbf{k}\right)  a_{2s\lambda}\left(
\mathbf{k}\right)  \right]  e^{ik\left(  x_{1}-x_{2}\right)  }.
\label{EAkspace}%
\end{align}
If $A_{2}=A_{1}$ the scalar product $\left(  A_{1},A_{2}\right)  _{t}$ reduces
to the spatial integral of number density $J^{0}\left(  t,\mathbf{x}\right)
$. Only free transverse photons are counted. In the presence of sources and
sinks this photon number is not conserved. The sine and cosine series are orthogonal.

Inspection of (\ref{EAcs}) and (\ref{BraKet}) shows that these expressions for
the scalar product involve both the potential and the electric field rather
than a single function. QM based on scalar products of this form can be
described within the formalism of biorthogonal QM \cite{Brody,HawtonDebierre}.
"Bi" refers to the use of a basis and its dual, while orthogonality refers the
fact that the scalar product is zero if the basis states are distinct. This
will be discussed in the next Section. The relationship between the notation
used here and that in \cite{MaxwellQM} is that here $a\left(  \mathbf{k}%
\right)  $ is a Fourier transform while in \cite{MaxwellQM} $c\left(
\mathbf{k}\right)  $ is the probability amplitude for a covariantly normalized
plane wave.

\section{Hilbert space and observables}

In this Section the Hilbert space and the momentum, position and angular
momentum operators and their eigenvectors are defined. Covariant normalization
that leads to expressions of the classical form is used. The probability
amplitude to find a photon at $\mathbf{x}$ on the $t$-hyperplane is calculated
and it is verified that the Born rule is satisfied. The even and odd fields
describing a physical state are assumed to be independent but the momentum and
position eigenvectors are bases that allow for both possibilities.

The real Hilbert space is the vector space of all $A_{c}$ and $A_{s}$ and
their derivatives with the scalar product (\ref{EAcs}). Momentum is an
observable. It can be verified by substitution in (\ref{EAkspace}) that the
plane waves with definite momentum $\hbar\mathbf{k}^{\prime}$ defined
covariantly as%
\begin{equation}
\mathbf{a}_{r\mathbf{k}^{\prime}\lambda^{\prime}}\left(  \mathbf{k}\right)
=\left(  2\pi\right)  ^{3}\omega_{k}\delta\left(  \mathbf{k}-\mathbf{k}%
^{\prime}\right)  \mathbf{e}_{\lambda^{\prime}}\left(  \mathbf{k}^{\prime
}\right)  \label{PlaneWaves}%
\end{equation}
with $\omega_{k}=c\left\vert \mathbf{k}\right\vert $ and $r=c$ and $s$ are
biorthogonal in the sense that%
\begin{equation}
\left(  A_{\mathbf{k\lambda}},A_{\mathbf{k}^{\prime}\lambda^{\prime}}\right)
=\delta_{\lambda\lambda^{\prime}}\left(  2\pi\right)  ^{3}\omega_{k}%
\delta\left(  \mathbf{k}-\mathbf{k}^{\prime}\right)  . \label{kBiorthogonal}%
\end{equation}
This normalization of the plane wave basis is invariant as can be seen from
$\int\frac{d\mathbf{k}}{\omega_{k}}\omega_{k}\delta\left(  \mathbf{k-k}%
^{\prime}\right)  =1$. In position space in the Heisenberg picture (HP)
\begin{equation}
\mathbf{A}_{\mathbf{k}\lambda}\left(  \mathbf{x},t\right)  =\sqrt{\frac{\hbar
}{\epsilon_{0}}}e^{-ikx}\mathbf{e}_{\lambda}\left(  \mathbf{k}\right)
\label{xPlaneWaves}%
\end{equation}
where $a_{c}=a_{s}$ so its real and imaginary parts are cosine and sine plane
waves. This contrasts with $\mathbf{A}$ describing a physical state in which
the cosine and sine terms should be independent real solutions to the wave
equation. In a general physical state $A$ the probability amplitude for wave
vector $\mathbf{k}$ is%
\begin{equation}
\left(  A_{\mathbf{k}\lambda},A_{r}\right)  =a_{r\lambda}\left(
\mathbf{k}\right)  \label{a(k)}%
\end{equation}
If $\omega_{k}$ is replaced with $\omega_{k}^{1/2}\ $in (\ref{PlaneWaves}) the
noncovariant Newton-Wigner \cite{NW} normalization, $\left(  A_{\mathbf{k}%
\lambda},A_{\mathbf{k}^{\prime}\lambda^{\prime}}\right)  _{NW}=\delta
_{\lambda\lambda^{\prime}}\left(  2\pi\right)  ^{3}\delta\left(
\mathbf{k}-\mathbf{k}^{\prime}\right)  $, is obtained. The position space
plane waves (\ref{xPlaneWaves}) are eigenvectors of the four-momentum operator%
\begin{equation}
\widehat{P}=\left(  \widehat{p}_{0},\widehat{\mathbf{p}}\right)  =\hbar\left(
\sqrt{-\nabla^{2}},-i\widehat{\epsilon}\mathbf{\nabla}\right)  \label{Pop}%
\end{equation}
In $\mathbf{k}$-space the four momentum operator is $\widehat{P}=\hbar\left(
\omega_{k}/c,\mathbf{k}\right)  $. In either case the four-momentum
eigenvalues are
\begin{equation}
P=\hbar\left(  \omega_{k}/c,\mathbf{k}\right)  . \label{Peigenvalues}%
\end{equation}
The Hamiltonian operator
\begin{equation}
\widehat{H}=\hbar c\sqrt{-\nabla^{2}} \label{H}%
\end{equation}
generates unitary transformations according to the Schr\"{o}dinger equation
\begin{equation}
i\hbar\widehat{\epsilon}\partial_{t}A\left(  t\right)  =\widehat{H}A\left(
t\right)  . \label{Schrodinger}%
\end{equation}

Position is also an observable. The Fourier transform of the localized state
$\delta\left(  \mathbf{x}-\mathbf{x}^{\prime}\right)  $ is the plane wave
$\exp\left(  i\mathbf{k\cdot x}^{\prime}\right)  $ so the photon position
eigenvectors in the HP should be of the form%
\begin{equation}
\mathbf{a}_{rx^{\prime}\lambda}\left(  \mathbf{k}\right)  =\omega_{k}^{\alpha
}\mathbf{e}_{\lambda}\left(  \mathbf{k}\right)  e^{ikx^{\prime}} \label{evecs}%
\end{equation}
for $r=c$ and $s$ where $kx^{\prime}=\omega_{k}t^{\prime}-\mathbf{k\cdot
x}^{\prime}$ and $\alpha=0$ in the covariant formulation and $\alpha=\frac
{1}{2}$ for Newton Wigner position eigenvectors. Here the covariance is
emphasized, so in the rest of this Section $\alpha$ will be set equal to $0$.
In position space%
\begin{equation}
\mathbf{A}_{\lambda}\left(  x\right)  =-\sqrt{\frac{\hbar}{2\epsilon_{0}}}%
\int_{t}\frac{d\mathbf{k}}{\left(  2\pi\right)  ^{3}\omega_{k}}\mathbf{e}%
_{\lambda}\left(  \mathbf{k}\right)  e^{-ik\left(  x-x^{\prime}\right)  }.
\label{Ax}%
\end{equation}
Using (\ref{EAkspace}) the projection of an arbitrary physical state described
by $A_{c}$ and $A_{s}$ onto the $A_{x\lambda}$ basis,%
\begin{equation}
\phi_{r\lambda}\left(  x\right)  =\left(  A_{x\lambda},A_{r}\right)  _{t}%
=\int_{t}\frac{d\mathbf{k}}{\left(  2\pi\right)  ^{3}\omega_{k}}a_{r\lambda
}\left(  \mathbf{k}\right)  e^{-ikx}, \label{phi}%
\end{equation}
has the mathematical form of a Lorentz invariant scalar potential that
satisfies the zero mass Klein-Gordon (KG) equation.

The Schr\"{o}dinger picture (SP) photon position operator with commuting
components and eigenvectors\ (\ref{evecs}) can be derived by rotating
$\mathbf{e}_{1}+i\lambda\mathbf{e}_{2}$ about $\mathbf{e}_{2}$ by $\theta$,
then about $\mathbf{e}_{3}$ by $\phi$ to give $\mathbf{e}_{\theta}%
+i\lambda\mathbf{e}_{\phi}$ using the operator $\widehat{R}$ so that
$\mathbf{e}_{\theta}+i\lambda\mathbf{e}_{\phi}=\widehat{R}\left(
\mathbf{e}_{1}+i\lambda\mathbf{e}_{2}\right)  $ and $i\mathbf{\partial
}_{\mathbf{k}}$ transforms to $\widehat{\mathbf{x}}=\widehat{R}%
i\mathbf{\partial}_{\mathbf{k}}\widehat{R}^{-1}$ \cite{HawtonBaylis}.
Alternatively it can be obtained by covariant differentiation \cite{Covariant}%
. It was originally obtained by brute force subtraction of the $\mathbf{k}%
$-space gradient of $\mathbf{e}_{\lambda_{j}}\left(  \mathbf{k}\right)  $ and
$\left\vert \mathbf{k}\right\vert ^{\alpha}$ to give \cite{HawtonPosOp}
\begin{equation}
\widehat{\mathbf{x}}=i\mathbf{\partial}_{\mathbf{k}}-i\alpha\frac{\mathbf{k}%
}{\left\vert \mathbf{k}\right\vert ^{2}}+\frac{1}{\left\vert \mathbf{k}%
\right\vert ^{2}}\mathbf{k\times}\widehat{\mathbf{S}}-\widehat{\lambda}%
\frac{\cos\theta}{k\sin\theta}\mathbf{e}_{\phi} \label{xop}%
\end{equation}
for position eigenvectors proportional to $\omega_{k}^{\alpha}$ where
$\omega_{k}=c\left\vert \mathbf{k}\right\vert $, the helicity operators is
$\widehat{\lambda}=\mathbf{e}_{\mathbf{k}}.\widehat{\mathbf{S}}$ for spin
operator $\widehat{\mathbf{S}}$ and the term $-i\mathbf{\partial}_{\mathbf{k}%
}\left\vert \mathbf{k}\right\vert ^{\alpha}=-i\mathbf{\partial}_{\mathbf{k}%
}\left\vert \mathbf{k}\right\vert ^{\alpha}=-i\left(  \frac{\alpha\mathbf{k}%
}{\left\vert \mathbf{k}\right\vert ^{2}}\right)  \left\vert \mathbf{k}%
\right\vert ^{\alpha}$ compensates for differentiation of $\omega_{k}^{\alpha
}$ in (\ref{evecs}). The HP photon position operator is given in
\cite{BabaeiMostafazadeh}. The helicity $\lambda$ photon position eigenvectors
have a definite component of total angular momentum in the fixed but arbitrary
direction $\mathbf{e}_{3}$ with indefinite spin and orbital contributions
\cite{HawtonBaylis}. The total angular momentum operator is
\begin{align}
\widehat{\mathbf{J}}  &  =\widehat{\mathbf{x}}\times\widehat{\mathbf{P}%
}+\widehat{\mathbf{J}}_{int},\label{Jtotal}\\
\widehat{\mathbf{J}}_{int}  &  =\hbar\lambda\left(  \frac{\cos\theta}%
{\sin\theta}\mathbf{e}_{\theta}\mathbf{+e}_{\mathbf{k}}\right)  . \label{Jint}%
\end{align}
where $\widehat{\mathbf{J}}_{int}$ is the internal angular momentum operator
and $\widehat{\mathbf{x}}\times\widehat{\mathbf{P}}$ describes external
angular momentum. The position and angular momentum operators are reviewed in
more detail in \cite{HawtonBaylis,MaxwellQM} where rotation about $\mathbf{k}$
through an Euler angle that gives a more general expression for $\mathbf{J}$
is included.

Setting $a_{r\lambda}\left(  \mathbf{k}\right)  $ in (\ref{phi}) equal to the
$\alpha=0$ physical states $a_{rx^{\prime}\lambda}\left(  \mathbf{k}\right)  $
given by (\ref{evecs}) with $\Delta t\equiv t-t^{\prime}$ and $r\equiv
\left\vert \mathbf{x}-\mathbf{x}^{\prime}\right\vert $ an explicit expression
for its time evolution can be obtained by taking sums and differences of%
\begin{align}
\int_{t}\frac{d\mathbf{k}}{\left(  2\pi\right)  ^{3}\omega_{k}}e^{-ik\left(
x-x^{\prime}\right)  } &  =\frac{1}{4\pi^{2}r}\sum_{\gamma=\pm}\left[
i\gamma\pi\delta\left(  r-\gamma c\Delta t\right)  \right.  \label{I}\\
&  \left.  +P\left(  \frac{1}{r-\gamma c\Delta t}\right)  \right]  \nonumber
\end{align}
to give%
\begin{align}
\phi_{c\lambda}\left(  x\right)   &  =\left(  A_{x\lambda},A_{cx^{\prime
}\lambda}\right)  \label{phic}\\
&  =\frac{1}{4\pi^{2}r}\left[  P\left(  \frac{1}{r-c\Delta t}\right)
+P\left(  \frac{1}{r+c\Delta t}\right)  \right]  ,\nonumber\\
\phi_{s\lambda}\left(  x\right)   &  =\left(  A_{x\lambda},A_{sx^{\prime
}\lambda}\right)  \label{phis}\\
&  =\frac{1}{4\pi r}\left[  \delta\left(  r+c\Delta t\right)  -\delta\left(
r-c\Delta t\right)  \right]  \nonumber
\end{align}
for $\lambda=\pm1$. Here $P$ denotes the principal value integral that
excludes the singularity in the integrand used to evaluate (\ref{I}). Both
(\ref{phic}) and (\ref{phis}) satisfy the homogeneous photon wave equation
$\square\phi_{r\lambda}\left(  x\right)  =0$, but only $\phi_{s\lambda}\left(
x\right)  $ is odd under QFT charge conjugation and couples to charged matter.
Schweber \cite{Schweber} inverted $\square$ and found that the unique Green's
function solving $\square\phi_{\lambda}\left(  x\right)  =\delta\left(
\mathbf{x}-\mathbf{x}^{\prime}\right)  \delta\left(  t-t^{\prime}\right)  $ is
$\frac{1}{4\pi r}\left[  \delta\left(  r+c\Delta t\right)  +\delta\left(
r-c\Delta t\right)  \right]  $ where $t^{\prime}=t-r/c<t$ is the retarded time
and $t^{\prime}=t+r/c>t$ is the advanced time. He concluded that the retarded
potential is determined by boundary conditions and equals the sum of his
unique contour integral independent Green function and a solution to a
homogeneous wave equation. This retarded potential is important in classical
EM and, according to (\ref{phis}), it can be applied to photon QM with the
significant advantage that $\phi_{s\lambda}\left(  x\right)  $ is a Lorentz scalar.

The scalar potential $\phi_{s\lambda}\left(  x\right)  $ is generally known as
the commutator or causal Green function since it has support only within the
light cone \cite{HalliwellOrtiz,ReehSchlieder}. Its time derivative
$\psi_{s\lambda}\left(  x\right)  \equiv i\partial_{t}\phi_{s\lambda}\left(
x\right)  $ evaluated at $t=t^{\prime}$,%
\begin{align}
\psi_{s\lambda}\left(  t,\mathbf{x}\right)   &  =\int_{t}\frac{d\mathbf{k}%
}{\left(  2\pi\right)  ^{3}}e^{-i\mathbf{k\cdot}\left(  \mathbf{x}%
-\mathbf{x}^{\prime}\right)  }\label{xbasis}\\
&  =\delta\left(  \mathbf{x}-\mathbf{x}^{\prime}\right)  ,\nonumber
\end{align}
forms a localized basis. The Born rule gives a probability interpretation of
the state vector. It states that if an observable corresponds to a
self-adjoint operator and the state vector describing a physical system is
normalized, the sum of the absolute squares of the probability amplitudes of
its eigenvalues is unity. For continuous observables such as position the sum
becomes an integral. Here as in the Schr\"{o}dinger description of the
electron the localized functions $\delta\left(  \mathbf{x-x}^{\prime}\right)
$ are not technically in the Hilbert space since they are not square
integrable but they satisfy a completeness relation and form a very convenient
basis \cite{CTQM1}, so the $\delta$-localized basis (\ref{xbasis}) will be
used here. Expanding $\psi_{s\lambda}$ in the $\delta$-basis at time $t$ as%
\begin{equation}
\psi_{s\lambda}\left(  t,\mathbf{x}\right)  =\int_{t}d\mathbf{x}^{\prime
}\delta\left(  \mathbf{x}-\mathbf{x}^{\prime}\right)  \psi_{s\lambda}\left(
t,\mathbf{x}^{\prime}\right)  , \label{psixbasis}%
\end{equation}
it can be seen that $\psi_{s\lambda}\left(  x\right)  $ is the probability
amplitude for a photon to be in the state $\delta\left(  \mathbf{x}%
-\mathbf{x}^{\prime}\right)  $ on the $t$-hyperplane. The $\lambda$-helicity
$\mathbf{x}$-space probability density is
\begin{equation}
\rho_{s\lambda}\left(  t,\mathbf{x}\right)  =\left[  \psi_{s\lambda}\left(
t,\mathbf{x}\right)  \right]  ^{2}. \label{photonxdensity}%
\end{equation}
If $\left(  A,A\right)  $ is finite, $A$ is normalizable as $\left(
A,A\right)  =1$. The $\mathbf{k}$-space probability density is%
\begin{equation}
\rho_{s\lambda}\left(  \mathbf{k}\right)  =\left[  a_{s\lambda}\left(
\mathbf{k}\right)  \right]  ^{2}. \label{kNorm2}%
\end{equation}
Quantum mechanics requires state vectors to describe physical systems and
operators representing observables such that the only possible result of a
measurement is one of their eigenvalues \cite{CTQM1}. Eqs. (\ref{xbasis}) to
(\ref{kNorm2}) provide a scalar Schr\"{o}dinger-like description of a photon
with helicity $\lambda$ in which $\psi_{s\lambda}\left(  t,\mathbf{x}\right)
$ is the probability amplitude for a photon to be at $\mathbf{x}$ on the
$t$-hyperplane and its Fourier transform $a_{s\lambda}\left(  \mathbf{k}%
\right)  $ is the probability amplitude for it to have momentum $\hbar
\mathbf{k}$.

\section{Second quantization}

A first quantized photon cannot be created or destroyed - creation and
annihilation of photons and the description of $n$-photon states requires
second quantization. In QFT, QED and Quantum Optics fields are second
quantized by raising them to the status of operators. The position
eigenvectors (\ref{Ax}) are positive frequency so these functions plus their
complex conjugates are real and become Hermitian operators when second
quantized. For an arbitrary first quantized state generalized to include a
factor $\omega_{k}^{\alpha}$ to accommodate the factor $\omega_{k}^{1/2}$
commonly used,
\begin{align}
\widehat{\mathbf{A}}\left(  x\right)   &  =\sqrt{\frac{\hbar}{\epsilon_{0}}%
}\sum_{\lambda=\pm1}\int_{t}\frac{d\mathbf{k}}{\left(  2\pi\right)  ^{3}%
}\left(  2\omega_{k}\right)  ^{\alpha-1}\left[  \widehat{a}_{\lambda}\left(
\mathbf{k}\right)  \mathbf{e}_{\lambda^{\prime}}\left(  \mathbf{k}\right)
e^{-ikx}\right. \nonumber\\
&  \left.  +\widehat{a}_{\lambda}^{\dagger}\left(  \mathbf{k}\right)
\mathbf{e}_{\lambda^{\prime}}^{\ast}\left(  \mathbf{k}\right)  \right]
e^{ikx}, \label{Aop}%
\end{align}
where the operator $\widehat{a}_{\lambda}\left(  \mathbf{k}\right)  $
annihilates a photon with wave vector $\mathbf{k}$ and helicity $\lambda$ and
$\widehat{a}_{\lambda}^{\dagger}\left(  \mathbf{k}\right)  $ creates one. The
plane waves will be assumed to satisfy the commutation relations
\begin{align}
\left[  \widehat{a}_{\lambda}\left(  \mathbf{k}\right)  ,\widehat{a}%
_{\lambda^{\prime}}\left(  \mathbf{k}^{\prime}\right)  \right]   &
=0,\ \left[  \widehat{a}_{\lambda}^{\dagger}\left(  \mathbf{k}\right)
,\widehat{a}_{\lambda^{\prime}}^{\dagger}\left(  \mathbf{k}^{\prime}\right)
\right]  =0,\nonumber\\
\left[  \widehat{a}_{\lambda}\left(  \mathbf{k}\right)  ,\widehat{a}%
_{\lambda^{\prime}}^{\dagger}\left(  \mathbf{k}^{\prime}\right)  \right]   &
=\delta_{\lambda,\lambda^{\prime}}\left(  2\pi\right)  ^{3}\left(  2\omega
_{k}\right)  ^{1-2\alpha}\delta\left(  \mathbf{k}-\mathbf{k}^{\prime}\right)
. \label{kcommutation}%
\end{align}
The usual text book choice is $\alpha=\frac{1}{2}$ but $\alpha=0$ for which
(\ref{Aop}) and (\ref{kcommutation}) are covariant is used here and in
\cite{ItzyksonZuber,VincentThesis}. The field operator (\ref{Aop}) creates and
annililates photons at $x$. The commutator%
\begin{align}
\widehat{C}_{\lambda}\left(  x,x^{\prime}\right)   &  \equiv i\frac
{\epsilon_{0}}{\hbar}\left[  \widehat{\mathbf{A}}_{\lambda}\left(
t,\mathbf{x}\right)  \cdot\widehat{\mathbf{E}}_{\lambda}\left(  t^{\prime
},\mathbf{x}^{\prime}\right)  \right. \nonumber\\
&  \left.  -\widehat{\mathbf{E}}_{\lambda}\left(  t^{\prime},\mathbf{x}%
^{\prime}\right)  \cdot\widehat{\mathbf{A}}_{\lambda}\left(  t,\mathbf{x}%
\right)  \right]  \label{C}%
\end{align}
with $\widehat{\mathbf{E}}_{\lambda}\left(  x\right)  =-\partial
_{t}\widehat{\mathbf{A}}_{\lambda}\left(  x\right)  $ describes creation of a
photon with helicity $\lambda$ at $\left(  t^{\prime},\mathbf{x}^{\prime
}\right)  $ followed by its annihilation at $\left(  t,\mathbf{x}\right)  $
and creation at $\left(  t,\mathbf{x}\right)  $ followed by annihilation at
$\left(  t^{\prime},\mathbf{x}^{\prime}\right)  $. It can be verified by
substitution at $t=t^{\prime}$ that%
\begin{equation}
\widehat{C}_{\lambda}\left(  t,\mathbf{x};t,\mathbf{x}^{\prime}\right)
=\delta\left(  \mathbf{x}-\mathbf{x}^{\prime}\right)  .
\end{equation}
Eq. (\ref{C}) leads to a physical interpretation of the potential
(\ref{phis}). Defining the vacuum state as $\left\vert 0\right\rangle $ and
the one photon position eigenvectors
\begin{equation}
\left\vert \mathbf{A}_{x\lambda}\right\rangle =\widehat{\mathbf{A}}_{\lambda
}\left(  x\right)  \left\vert 0\right\rangle ,\ \left\vert \mathbf{E}%
_{x\lambda}\right\rangle =\widehat{\mathbf{E}}_{\lambda}\left(  x\right)
\left\vert 0\right\rangle \label{OnePhoton}%
\end{equation}
and using the standard bra-ket notation, the vacuum expectation value of
(\ref{C}) gives
\begin{equation}
\left\langle 0\left\vert \widehat{C}_{\lambda}\left(  x,x^{\prime}\right)
\right\vert 0\right\rangle =\phi_{s\lambda}\left(  x\right)
\label{VacuumExpectation}%
\end{equation}
that propagates causally.

\section{Summary and Conclusion{}}

In its covariant $\alpha=0$ version, photon quantum mechanics as described
here preserves the classical form of the EM potential and fields when first
and second quantized. Only the interpretation need be changed - from real
observable classical fields, to probability amplitudes, and then to operators
that create and annihilate photons. The real potentials are even and odd under
QFT charge conjugation, but only those that are odd can be localized in a
finite region and coupled to charged matter. These even and odd fields are
real and imaginary parts of a complex field whose use simplifies the
mathematics and facilitates use of the standard Lagrangian and relativistic
scalar product. Propagation of finite pulses is as in classical EM and
mathematical techniques such as finite difference time domain (FDTD)
\cite{FDTD} developed to handle problems in classical EM theory can be applied
directly to single photons. Projection onto momentum and position bases gives
a covariant Schr\"{o}dinger-like description of photon QM. Eqs. (\ref{phi})
and (\ref{psixbasis}) to (\ref{kNorm2}) provide a scalar description of single
photon states with a well defined physical interpretation that may prove to be
useful in applications.

The need for a single photon wave function consistent with classical EM theory
is illustrated by the interpretation of a recent experiment. Propagating light
pulses were split by a Fresnel biprism and coincidence counts were registered
\cite{LightQuantum}. For a single photon emitted by a color center in a
diamond nanocrystal no unexplained coincidences were observed, while for faint
laser pulses there were ten times more coincidence counts.\ This is a clear
demonstration that a one-photon state exhibits quantum mechanical
particle-like behavior. In analysis of their data the authors use "the
well-know result from Quantum Optics that phenomena like interference,
diffraction, propagation, can be computed with the classical theory of light
even in the single-photon regime."

\end{document}